\documentstyle[12pt]{article}
\def\hybrid{\topmargin -20pt    \oddsidemargin 0pt
        \headheight 0pt \headsep 0pt
       \textwidth 6.5in        % US paper
       \textheight 9in         % US paper
        \marginparwidth .875in
        \parskip 5pt plus 1pt   \jot = 1.5ex}

\hybrid

\newcommand{\hf}{\frac12}
\newcommand{\qt}{\frac14}
\newcommand{\bea}{\begin{eqnarray}}
\newcommand{\eea}{\end{eqnarray}}
\newcommand{\be}{\begin{equation}}
\newcommand{\ee}{\end{equation}}
\newcommand{\bt}{\begin{tabular}}
\newcommand{\et}{\end{tabular}}
\newcommand{\ba}{\begin{array}}
\newcommand{\ea}{\end{array}}

\newcommand{\vev}[1]{\langle #1 \rangle}
\newcommand{\tr}{{\rm Tr}}
\newcommand{\pf}{{\rm Pf}}

\newcommand{\drawsquare}[2]{\hbox{%
\rule{#2pt}{#1pt}\hskip-#2pt%  left vertical
\rule{#1pt}{#2pt}\hskip-#1pt%  lower horizontal
\rule[#1pt]{#1pt}{#2pt}}\rule[#1pt]{#2pt}{#2pt}\hskip-#2pt%  upper horizontal
\rule{#2pt}{#1pt}}% right vertical
\newcommand{\fun}{\raisebox{-.5pt}{\drawsquare{6.5}{0.4}}}

\newcommand{\as}{\raisebox{-3.5pt}{\drawsquare{6.5}{0.4}}\hskip-6.9pt%
        \raisebox{3pt}{\drawsquare{6.5}{0.4}}}

\def \half{{\textstyle\hf}}
\def \quarter{{\textstyle\qt}}
\def \nonpert{non\discretionary{-}{}{-}per\-tur\-bative\ }
\def\Gpert{G_{\rm pert}}
\def\WNP{W^{\rm (NP)}}
\def\fNP{f^{\rm (NP)}}
\def\GNP{G_{\rm NP}}
\def\Mpl{M_{\rm Pl}}
\begin{document}
\begin{titlepage}
\begin{center}

\hfill hep-th/9707212\\

\vskip .6in
{\large \bf 
Singular Yukawa and gauge couplings\\
in $d=4$ Heterotic String Vacua}\footnote{Research 
 supported by: the DFG (M.K.);
 the NATO, under grant CRG~931380 (J.L.);
 GIF -- the German--Israeli
Foundation for Scientific Research (J.L.).}
\vskip .5in

{\bf Matthias Klein$^{a,b}$ and  Jan Louis$^{a}$}
\footnote{email: mklein@hera2.physik.uni-halle.de, 
jlouis@hermes.physik.uni-halle.de
}
\\
\vskip 0.8cm
{}$^{a}${\em Martin--Luther--Universit\"at 
Halle--Wittenberg,\\
        FB Physik, D-06099 Halle, Germany}
\vskip 0.5cm
{}$^{b}${\em Sektion Physik, 
Universit\"at M\"unchen,\\
Theresienstr.~37, D-80333 M\"unchen, Germany}

\end{center}

\vskip 1.5cm

\begin{center} {\bf ABSTRACT } \end{center}
In this paper we discuss
the singularities in the Yukawa and 
gauge couplings of $N=1$ compactifications of 
the $SO(32)$ heterotic string in four space-time
dimensions. Such singularities can arise from
the strong coupling dynamics of a confined
non-perturbative gauge group.

\vfill
July 1997
\end{titlepage}

\def\baselinestretch{1.2}
\baselineskip 16 pt

\section{Introduction}
The recent advances in string theory have led 
to a much better understanding of its
non-perturbative properties.
One particular aspect of these developments
concerns couplings
in the low energy effective action,
which are singular functions on the 
moduli space of a given family of perturbative 
string vacua.
Whenever such couplings
obey a non-renormalization theorem,
the singularity cannot be smoothed
out by (perturbative) 
quantum corrections but must have 
a physical origin.
An example of this situation is found
in $K3$ compactifications of the heterotic 
string which have six space-time dimensions 
($d=6$) and 8 supercharges.
For consistency, such vacua necessarily 
have  non-trivial
instantons embedded in
the gauge group 
$E_8\times E_8$ or $SO(32)$.
Associated with these instantons is a 
(quaternionic) moduli space 
which parametrizes the size, location, 
and embedding of the instantons
into the gauge group.
This moduli space has singularities at points
where instantons shrink to zero size.
It has been convincingly argued that
the physical origin of these singularities 
is either a set of gauge bosons becoming
massless or an entire string becoming tensionless
[\ref{schmal}-\ref{AM}].
%\cite{schmal,hanany,SW,bele,aspinwall,intriligator,AM}.
Both of these effects are invisible in string perturbation theory
in that they occur in regions of the moduli space
where the perturbation theory breaks down.

A similar situation is found in 
type II vacua compactified on 
Calabi--Yau threefolds 
which have $N=2$ supersymmetry in $d=4$.
In this case the gauge couplings generically
have logarithmic singularities
on the moduli space of the scalar fields
in the $N=2$ vector multiplets.
Strominger \cite{strominger}
suggested that these singularities 
are caused by \nonpert charged 
states in the type II 
string spectrum which become massless
at the singularity. 
In perturbative string theory all such states 
are heavy and hence are integrated out 
of the effective action. Their only remnant
visible in perturbation theory are
the moduli dependent (and singular) 
gauge couplings.
However, whenever some of 
the \nonpert states become massless 
it is no longer legitimate to integrate 
them out, and thus the singularity signals  
additional light states
and the break down of an  effective low energy
theory which does not properly include all 
the light degrees of freedom.
The non-perturbative states arise in $N=2$
hypermultiplets and 
carry $U(1)$ charge of Ramond--Ramond vector bosons.
As a consequence, they do not have the canonical
couplings to the dilaton \cite{various}
but their mass only depends on
the scalar fields of the $N=2$ 
vector multiplets.
Therefore, the corrections to the gauge couplings 
-- even though being non-perturbative --
appear without
any dilaton dependence and thus `compete' with
tree level effects.

In $N=1$ heterotic string vacua
compactified on a Calabi--Yau manifold
or, more generally, on $(0,2)$ 
superconformal field theories (SCFTs) 
the situation is
more complicated. In this case
one typically has power-like singularities
of the tree level Yukawa couplings
in addition to logarithmic singularities
of the gauge couplings at one-loop.
The power-like tree level singularities
of the Yukawa couplings
cannot be explained by states becoming
massless. Instead it was suggested
by Kachru, Seiberg and Silverstein \cite{KSS} 
that at least some of the singularities
are caused by strong
coupling dynamics of an asymptotically free
\nonpert gauge group.
As we already stated, \nonpert gauge groups
do arise in $d=6$ vacua of the heterotic string.
For example,
$Sp(2k)$ appears when $k$ small $SO(32)$
instantons
shrink on the same smooth point in the $K3$ \cite{schmal}
while $U(2k)$ is the \nonpert gauge group 
for $k$ small instantons (without vector structure)
shrinking
on a singular point of $K3$ 
[\ref{BS}-\ref{DM},\ref{bele}-\ref{AM}].
%\cite{BS,GP,DM,bele,aspinwall,intriligator,AM}.
It has been 
shown  
that for a specific class of Calabi--Yau 
compactifications the \nonpert gauge group
`descends down' to  $d=4$ and can be responsible
for the singular couplings of the effective action \cite{KSS}.

In this paper we expand on 
the mechanism of ref.~\cite{KSS} 
in two respects. 
We show that generically there can be
states in the string spectrum 
which become massless at the singularity
and thus are responsible for the 
additional logarithmic
singularity of the gauge couplings at 
one-loop (section~2). Furthermore, ref.~\cite{KSS} 
concentrated on a one-dimensional moduli
space and a \nonpert gauge group $SU(2)$.
The generalization to higher dimensional
moduli spaces with a more complicated 
structure of the singularities and \nonpert gauge groups
$Sp(2k)$ and $U(2k)$ 
are presented in sections~3 and 4 respectively.

A summary of our results appeared previously
in ref.~\cite{klein}.

\section{Heterotic vacua with non-perturbative
gauge group $SU(2)$}

Perturbative $d=4$, $N=1$ heterotic vacua
are characterized by a $c=9$ SCFT
with $(0,2)$ worldsheet supersymmetry
and the choice of a $\bar c = 22$
vector bundle.\footnote{An important subset
of such vacua are the geometrical Calabi--Yau
compactifications which have $(2,2)$ 
worldsheet supersymmetry.}
 The space-time spectrum features
the gravitational multiplet, non-Abelian vector
multiplets, charged chiral matter multiplets $Q^I, \hat Q^{\hat I}$,
and gauge neutral chiral moduli multiplets $M^i$.
The couplings of these multiplets
are described by an effective Lagrangian which is 
constrained by $N=1$ supersymmetry to only depend 
on three arbitrary functions: 
the real K\"ahler potential $K$, 
the holomorphic superpotential $W$, and the 
holomorphic gauge kinetic function $f$.
Due to their holomorphicity the latter two
obey a non-renormalization theorem \cite{LF}.
The superpotential $W$ receives no perturbative
corrections, and one only has
\be
W=W^{(0)} + \WNP\ ,
\ee
where $W^{(0)}$ denotes the tree level contribution
while $\WNP$ summarizes the \nonpert 
corrections. 
$W^{(0)}$ contains mass terms and 
Yukawa couplings both of which are 
generically moduli dependent. 
$\hat Q^{\hat I}$ are massive charged multiplets
of the string vacuum while $Q^I$ denotes the massless multiplets.
The massive modes $\hat Q^{\hat I}$ are 
commonly integrated out of
the effective Lagrangian since their typical mass is of order
of the Planck scale $\Mpl$. However, since their masses can be 
moduli dependent, they might become light in special regions
of the moduli space, and therefore we choose
to keep them in the effective theory. Thus a generic 
tree level superpotential is given by
\be
W^{(0)} = 
m_{\hat I}(M^i)\, \hat Q^{\hat I} \hat Q^{\hat I}
+ 
Y_{IJK}(M^i)\, Q^I  Q^J  Q^K + \ldots \ ,
\ee
where all gauge quantum numbers of 
$Q^I$ and $\hat Q^{\hat I}$ are suppressed.

The real part of the gauge kinetic function  $f$ determines the 
inverse gauge coupling according to
$g^{-2} = {\rm Re} f$.
The holomorphic $f$ receives no perturbative
corrections beyond one-loop, and one has
\be
f=f^{(0)}+f^{(1)}  + \fNP\ ,
\ee
where $f^{(1)}$ is the one-loop correction.

In most (if not all) heterotic string vacua
$W$ and $f$ are singular functions on the 
moduli space. (For a recent discussion
see ref.~\cite{KS}.)
For example, in a heterotic vacuum
obtained as a Calabi--Yau compactification
on  a quintic hypersurface in ${\bf CP^4}$ 
with  defining polynomial 
$p=\sum_{\alpha=1}^{5} X_\alpha^5 
- 5\psi X_1X_2X_3X_4X_5$
one finds \cite{CDGP,BCOV}
\be
Y \sim M^{-1} , \quad f \sim \log M\ ,
\ee
where $M\equiv 1-\psi^5$.
It has been 
suggested in ref.~\cite{KSS} that such singularities
are consequences of a strongly
coupled  \nonpert gauge group.
Such \nonpert gauge groups are known to 
arise in the $d=6$ $SO(32)$ heterotic string 
at points where instantons shrink to zero size.
The moduli space  of $k$ $SO(32)$ instantons
at smooth points of $K3$
is isomorphic to the Higgs branch
of an $Sp(2k)$\footnote{By $Sp(2k)$\ we mean the rank $k$\ 
symplectic group whose fundamental representation has 
dimension $2k$.} gauge theory with 32 half-hypermultiplets
 in the fundamental 
($\fun$) $2k$-dimensional representation,
a traceless antisymmetric tensor ($\as$) in the
$k(2k-1)-1$-dimensional representation, and 
a singlet \cite{schmal}. 
The singularities of this moduli space
precisely occur where some (or all) of the 
$Sp(2k)$ gauge bosons become massless.\footnote{
Singularities in the
$E_8\times E_8$ heterotic string 
and chirality changing phase transitions
have recently been discussed
in refs.~\cite{K-S,ibanez}.}
Upon toroidally compactifying the above theories one obtains
$N=2$ string vacua in $d=4$. String vacua with $N=1$ 
supersymmetry arise when one compactifies not on $K3\times T^2$
but on a Calabi--Yau threefold. However, there is a particular class
of threefolds -- $K3$ fibrations --
which is closely related to the six-dimensional
heterotic vacua. For such vacua a $K3$ is fibred
over a ${\bf P^1}$ base, and if the base is large 
the adiabatic argument
applies \cite{VW}, and the singularities
of the $K3$ fibres are inherited from the corresponding
six-dimensional vacuum \cite{KSS}. 
In this case the gauge group $G$ of the 
four-dimensional string vacuum
is a product of the perturbative
gauge group $\Gpert$ and the \nonpert gauge group
$\GNP$
\be
G=\Gpert\times\GNP\ ,
\ee
where $\GNP$ is a subgroup of $Sp(2k)$.

A specific class of $K3$ fibrations
are the quintic hypersurfaces 
defined in weighted projective space
${\bf WP^4}_{1,1,2k_1,2k_2,2k_3}$ \cite{CDFKM,KLM}.
For compactifications of the $SO(32)$ heterotic string on
$(0,2)$ deformations of
such Calabi--Yau spaces the \nonpert 
spectrum is computed in ref.~\cite{KSS}
for the case of a single small instanton
at a smooth point in the $K3$ fibre.\footnote{It is important 
to consider a $(0,2)$ deformation since on the $(2,2)$ locus the spin
connection is embedded in the gauge connection, and a small
instanton necessarily has to shrink on a $K3$ singularity. 
This situation is not fully understood at present.
(We thank P.~Aspinwall and K.~Intriligator
for a useful correspondence on this point.)}
It is found that the \nonpert gauge group in the four-dimensional
vacuum is given by 
$\GNP = Sp(2) \cong SU(2)$, and out of the 32 half-hypermultiplets
in $d=6$ 
only four $SU(2)$ doublets in chiral $N=1$ supermultiplets 
(which we denote by $q_i^\alpha, i=1, \ldots, 4,
\alpha=1,2$)
survive in $d=4$. 
This resulting gauge theory is asymptotically free 
($b_{SU(2)}>0$) and thus becomes
strongly coupled below its characteristic scale $\Lambda_{SU(2)}$.
It has the additional property that 
\be\label{cdef}
c:= T({\bf ad}) - \sum_r n_r T(r) 
= 2-4\cdot \frac12 =0\ ,
\ee
where $T(r)$ is the index in the representation $r$,
$T({\bf ad})$ is the index in the adjoint representation, and
$n_r$ counts the number of chiral multiplets in 
representation $r$. 
(In this notation the one-loop coefficient
of the $\beta$-function is given by
$b=3T({\bf ad}) - \sum_r n_r T(r)$.)

The coefficient $c$ also appears in the anomaly 
equation of the R-symmetry. An anomaly free
R-symmetry imposes
$\sum_r n_rT(r)\, R_r = -c$,\ where
$R_r$\ is  the R-charge of a  superfield
in representation $r$.
Therefore, in gauge theories with $c=0$ 
one can always choose $R=0$ for all superfields,
and, as a consequence,
no \nonpert superpotential $\WNP$ (which 
necessarily carries $R=2$)
can be generated by the strong coupling 
dynamics \cite{ADS,confine}. This conclusion
is believed to hold irrespective of the precise
form of the tree level superpotential.

The $SU(2)$ gauge theory under consideration  
confines below $\Lambda_{SU(2)}$, 
and the surviving
degrees of freedom are the gauge singlets 
\be
M_{ij} := q^\alpha_i \epsilon_{\alpha\beta} q^\beta_j\ .
\ee
$M_{ij}$ is antisymmetric, and due to Bose statistics of the
quark superfields it obeys the constraint
$\pf M:= {1\over 8}\epsilon^{ijkl} M_{ij} M_{kl} =0$; thus 
there are five physical degrees of freedom
in the effective theory.
Quantum mechanically the constraint
is modified and reads \cite{Seib_D49}
\be \label{su2_constr}
\pf M= 
\Lambda_{SU(2)}^4 \ .
\ee

Non-renormalizable interactions typically induce
mass terms as well as higher dimensional couplings
for some of the $M_{ij}$, and thus 
not all scalar degrees of freedom are moduli
of the low energy theory. 
Altogether the superpotential is given by 
\bea\label{superp}
W = m_{\hat I} (M)\, \hat Q^{\hat I} \hat Q^{\hat I}
+ Y_{IJK}(M)\, Q^I Q^J Q^K 
+ \lambda (\pf M-\Lambda^4) 
+ \sum_{i<j} m_{ij}M_{ij}^{2}\nonumber\\
+\ \hbox{non-renormalizable terms}\ 
+ \hbox{stringy non-perturbative terms}\ ,
\eea
where $\lambda$ is a Lagrange multiplier and
we have choosen $\Mpl=1$.
Note that even though no non-perturbative 
superpotential is generated by the strongly coupled
gauge theory, there are `stringy'
non-perturbative corrections of order
${\cal O}(e^{-g_{\rm string}^{-2}})$
where $g_{\rm string}$ is the string coupling 
constant which is related to vacuum expectation
value of the dilaton.

The supersymmetric minima
are determined by the solution of $\partial W =0$,
where the derivative has to be taken with respect to
all fields in the theory. The matter fields
$Q^I,\hat Q^{\hat I}$ of a string vacuum
are always choosen to obey
$\vev{Q^I}=\vev{\hat Q^{\hat I}}=0$ which leaves us
with
\bea \label{eom_mass}
{\partial W\over\partial M_{ij}}=0 &\to 
  &\half\lambda\ \epsilon^{ijkl}M_{kl}+2m_{ij}M_{ij}=0\ ,\\
\label{eom_cons}
{\partial W\over\partial\lambda}=0 &\to 
  &\epsilon^{ijkl}M_{ij}M_{kl}=8\ \Lambda^4
\eea
as nontrivial conditions.
(We neglect further non-renormalizable interactions,
since they are suppressed by $\Mpl$.
In addition, $g_{\rm string}$ is taken to be weak
so that the stringy 
non-perturbative corrections can also be ignored.)

To further simplify our analysis we only consider
solutions of eqs.~(\ref{eom_mass}), (\ref{eom_cons})
that satisfy $\vev W =0$.
This is partly motivated by the fact that 
$\vev W \neq 0$ induces a cosmological constant
in supergravity and partly by the considerable
simplification of the vacuum solutions.
Inserting eqs.~(\ref{eom_mass}), (\ref{eom_cons})
into eq.~(\ref{superp}) and using 
$\vev{Q^I}=\vev{\hat Q^{\hat I}}=0$
we arrive at 
\be
\vev W =
\sum_{i<j}m_{ij}\vev{M_{ij}}^2=
-\vev\lambda\Lambda^4\ .
\ee
Thus $\vev W =0$ demands $\vev\lambda=0$.
This class of solutions has an immediate other
consequence. From eq.~(\ref{eom_mass}) we learn
that each of the $M_{ij}$ whose mass term 
does not vanish is set to zero. 
Eq.~(\ref{eom_cons}) then implies that
at least
two masses have to vanish, 
and they have to be such that $m_{ij}=m_{kl}=0$
with  $i,j,k,l$ all different. 
Thus, in our approximation the dimension of the moduli
space is 
given by the number of vanishing mass terms
minus one.\footnote{Strictly speaking
the condition 
for a given $M_{ij}$ to be a modulus
involves higher dimensional terms
which we have neglected in our analysis.}

Ref.~\cite{KSS} considered the choice
$m_{12}=m_{34}=0$ which 
implies a one-dimensional
moduli space and 
\be\label{sol_su2}
M_{13}= M_{14}=M_{23}=M_{24} = 0 \ , \qquad
M_{12}M_{34} = \Lambda^4\ ,
\ee
via eqs.~(\ref{eom_mass}), (\ref{eom_cons}).
(We suppress the
brackets to denote VEV's henceforth.)

In string perturbation theory both 
the Yukawa couplings $Y_{IJK}(M)$ and the masses 
$m_{\hat I} (M)$  are given as power series expansion
in the moduli. However, the strong coupling effects which are responsible
for generating the \nonpert constraint
$\pf M= \Lambda_{SU(2)}^4$ remove the origin of the moduli space
and render the perturbative expansion of the Yukawa couplings
singular \cite{KSS}. For example,
\be\label{Ytree}
Y_{IJK}(M) \sim M_{12}+ M_{34} + \ldots
\ee
produces a singularity 
\be\label{singularity}
Y_{IJK}(M) \sim \frac{\Lambda^4}{M_{34}} 
+ \ldots \quad {\rm as\ }M_{34}\to 0\ .
\ee
(Higher dimensional terms in (\ref{Ytree}) 
are Planck suppressed but do produce additional
singular terms in eq.~(\ref{singularity}).) 

The $SU(2)$ scale 
$\Lambda_{SU(2)}$ 
depends on the gauge coupling $g_{SU(2)}$  via 
$\Lambda_{SU(2)} \sim {\rm exp}({-\frac{8\pi^2}{bg^2_{SU(2)}}})$.
It was shown in refs.~\cite{sagnotti,DMW} that 
in $d=6$ the gauge couplings of the \nonpert
gauge bosons do not depend on $g_{\rm string}$ 
or, in other words, do not have the canonical
couplings to the dilaton. It can be easily seen
that the same properties hold in $d=4$
and hence 
the Yukawa couplings in eq.~(\ref{singularity})
do not depend on $g_{\rm string}$ either. 
Thus, exactly as for the singularities of the
 $N=2$ vacua,
the non-perturbative effect which generates the 
singularity in eq.~(\ref{singularity})
does not have the standard dilaton or $g_{\rm string}$
dependence but instead competes with tree level
couplings. 
(Of course, this is a necessary 
requirement for a consistent
explanation of the singularities in the Yukawa couplings.)

The mass terms  $m_{\hat I}$  either become large
or small at the singular points in the moduli space. 
We already learned from the $N=2$ case \cite{strominger}
that light matter fields  $\hat Q^{\hat I}$
with $m_{\hat I} (M) \sim M_{34}$ 
 which are charged under 
$\Gpert$ produce a singularity
in the associated (perturbative) 
gauge coupling at one-loop;
more precisely they contribute a correction 
\be\label{FTsing}
g_{\rm pert}^{-2}= 
\sum_{\hat I} \frac{b_{\hat I}}{16\pi^2}\,
 {\rm log} |m_{\hat I}|^2 
+\ldots\ .
\ee
The coefficient of the singularity is set 
by the multiplicity
of the light modes and their gauge quantum numbers.
Unfortunately, in string theory the
coefficient of the logarithmic singularity
is so far only known for
$(2,2)$ vacua \cite{BCOV,KL}, and hence a more detailed comparison with the 
$(0,2)$ models considered in ref.~\cite{KSS} cannot be presented.
Conversely, it is not known at present how to repeat the 
analysis of ref.~\cite{KSS} for $(2,2)$ vacua, since the 
\nonpert physics in the corresponding 
$d=6$ vacua is not completely understood.
Nevertheless it is interesting to display the 
coefficients of the singularity in 
$(2,2)$ vacua.\footnote{This was worked out jointly
with V.~Kaplunovsky.}
One finds for  all 
$(2,2)$ vacua of the $SO(32)$ heterotic string, where
$\Gpert = SO(26)\times U(1)$, the relation
\be\label{diff_so}
16\pi^2 \big(g^{-2}_{SO(26)} 
- {\textstyle\frac{1}{6}} g^{-2}_{U(1)}\big) = -16 F_1\ ,
\ee
while for $(2,2)$ vacua of the $E_8\times E_8$
heterotic string ($\Gpert = E_8\times E_6$)
one has
\be\label{diff_e}
16\pi^2 \big(g^{-2}_{E_8} - g^{-2}_{E_6}\big) = -12 F_1\ .
\ee
$F_1$ is the topological index defined in 
ref.~\cite{BCOV} which for the quintic hypersurface
in ${\bf CP^4}$ is given by
\be\label{Fone}
F_1 = - \frac{1}{12}\, {\rm log} M \bar M + \ldots \ .
\ee
One can check that   
eqs.~(\ref{FTsing})--(\ref{Fone}) are not easy
to satisfy.
Let us first consider a pair of 
light states in the fundamental representation of 
$SO(26)$ with $U(1)$ charges $\pm q$.
Inserting eq.~(\ref{FTsing}) into (\ref{diff_so})
 one obtains 
\be
(-2+\frac{52}6\, q^2)\ {\rm log} |m|^2 = -16 F_1 =
\frac43\, {\rm log} |M|^2\ ,
\ee 
where the last equation used (\ref{Fone}).
Already for $q=1$ this implies the relation
$m = M^{\frac1{5}}$. However, the mass of 
a state given by a fractional power
of $M$ is not sensible within our framework.
Similarly, there is no sensible solution for
fields in two-index tensor or spinor representations
of $SO(26)$.
Thus we conclude that at the singularity there can 
be no light states which are charged under $SO(26)$.
For a pair of $SO(26)$ singlets with $U(1)$ charges 
$\pm q$ one finds instead
\be
\frac13\, q^2\, {\rm log} |m|^2 = -16 F_1 =
\frac43\, {\rm log} |M|^2\ .
\ee 
This relation can be satisfied 
by one pair with mass $m=M$ and $U(1)$ charges $\pm 2$
or four pairs with mass $m=M$ and $U(1)$ charges $\pm 1$.
(If  $m=M^2$ ($m=M^4$) one can also have two 
(one) pairs with $U(1)$ charges $\pm1$.)
The case of 
$E_8\times E_6$ was already discussed in ref.~\cite{KL}, 
where it was shown that no sensible solution 
exists at all.
The difficulty in satisfying 
eqs.~(\ref{FTsing})--(\ref{Fone}) 
might indicate that in
$(2,2)$ vacua a different mechanism 
is responsible for the singularities 
in the gauge couplings as well as in the Yukawa
couplings.\footnote{We checked that this conclusion
also holds for the singularities in the $K3$ fibre
of the two-parameter models of 
ref.~\cite{CDFKM} and we suspect that it is 
valid in general, since the coefficient of the
conifold singularities seems to be universal.}

So far we considered a one-dimensional
moduli space with 
$m_{12}=m_{34}=0$. By choosing additional
mass terms to be zero, one can obtain higher 
dimensional moduli spaces. However, the 
singularity of the Yukawa couplings
remains essentially unchanged
in that it is a 
smooth function of the additional moduli.
This follows immediately from 
the quantum constraint (\ref{eom_cons}) which 
is only quadratic in the fields.
In order to encounter a more complicated singular
structure the quantum constraint has to involve
higher powers in the fields. This precisely
occurs in $Sp(2k)$ gauge theories which we turn to
in the following section.

%%%%%%%%%%%%%%%%%%%%%%%%%%%%%%%%%%%%%
\section{Heterotic vacua with non-perturbative
gauge group $Sp(2k)$}

In most Calabi--Yau vacua the singularity of the 
Yukawa couplings has a more complicated structure 
than just a simple pole.
In particular, one observes generically a singular locus
with more than one component where the different
components can intersect in various ways.
Such a behaviour is reproduced by $k$ small instantons
located at the same (smooth) point in moduli space or, in other words,
by $\GNP=Sp(2k)$. For this case the analysis of ref.~\cite{KSS} 
can be repeated without major modifications, since it does not 
depend on the non-perturbative gauge group.
Thus, out of the 32 half-hypermultiplets 
in the fundamental ($\fun$) representation 
of the six-dimensional vacua again four chiral $\fun$ 
multiplets ($q_i^\alpha,i=1,\ldots,4,\alpha=1,\ldots,2k$)
remain in $d=4$. In addition, the gauge bosons 
as well as the antisymmetric tensor ($\as$) are constants
on the ${\bf P}^1$ base of the $K3$ fibration
and hence they also 
survive as chiral multiplets.
As before, the resulting spectrum in $d=4$ is 
an asymptotically free
gauge theory with the additional property that 
$c_{Sp(2k)} = (k+1) - 4\cdot\frac12 - (k-1)= 0$
for all values of $k$. Thus, the R-charge 
of all fields can be choosen to vanish 
and no non-perturbative 
superpotential can be generated by 
the strongly coupled $Sp(2k)$ gauge theory.
Fortunately, this gauge theory has been analysed in some detail 
in ref.~\cite{CSS} (see also \cite{CK,IP}). 
The physical degrees of freedom 
below $\Lambda_{Sp(2k)}$ are found to be
\bea\label{MTdef}
M_{ij}^l &:= &q_i \cdot A^l \cdot q_j\ =\ q_i^\alpha A_\alpha^{\ \beta_1}
A_{\beta_1}^{\ \beta_2}\cdots A_{\beta_{l-1}}^{\ \gamma}q_{j\gamma},
\qquad i,j=1,\ldots,4,\quad l=0,\ldots,k-1\ ,\nonumber\\
T_r &:= &\frac{1}{4r}\,\tr A^r\ =\ \frac{1}{4r}\, A_\alpha^{\ \beta_1}
A_{\beta_1}^{\ \beta_2}\cdots A_{\beta_{r-1}}^{\ \alpha}
,\qquad r=2,\ldots,k\ ,
\eea
where $A$ denotes the antisymmetric tensor, and colour indices are
lowered using the $Sp(2k)$ invariant $J$-tensor, 
$A_\alpha^{\ \beta}=J_{\alpha\gamma}A^{\gamma\beta}$.
Exactly as for the $SU(2)$ gauge theory the $M_{ij}^l$
and $T_r$ are not all independent but related by constraint equations.
These constraints can be obtained from the confining superpotential 
$W_{F=6}$ of the theory with  six \fun\ fields ($F=6$)
by integrating out two $q$'s \cite{CSS}. 
$W_{F=6}$ is uniquely determined by symmetry
arguments and the requirement
that the equations of motion reproduce the
classical constraints \cite{ADS,CSS}.
One obtains 
\be \label{W_conf}
W_{F=6}={1\over\Lambda^{2k+1}}\sum_{l,m,n,\{i_r\}} c_{lmn,i_2\ldots i_k} 
           \prod_{r=2}^k (T_r)^{i_r} M^l M^m M^n\ ,
\ee
where the sum runs over all integers 
\bea
&0\le l,m,n\le k-1\ ,\quad i_r\ge 0\ ,\nonumber\\
&{\rm with}\quad\displaystyle l+m+n+\sum_{r=2}^k r\, i_r=2(k-1)\ ,
\eea
and the flavour indices of the three $M$'s 
in eq.~(\ref{W_conf}) 
are contracted with an 6-index
epsilon tensor. Some of the coefficients in (\ref{W_conf}) may vanish;
for $k\le 4$ they are calculated in \cite{CSS}. 
Adding a mass term
for two of the quarks (e.g., $W_{mass}=mM^0_{56}$) and integrating out
the massive modes, one obtains $k$ constraints on the $F=4$ moduli space.
These constraints are in one-to-one correspondence with algebraic relations 
which already hold at the classical level among the
fields. Only one of these relations receives quantum corrections;
it has the form
\be \label{Sp_constr}
\sum_{l,m,\{i_r\}} \tilde c_{lm,i_2\ldots i_k} \prod_{r=2}^k (T_r)^{i_r} 
      M^l M^m =\Lambda^{2k+2}\ ,
\ee
with
\be \label{Sp_cond}
l+m+\sum_{r=2}^k r\, i_r=2(k-1)\ .
\ee
The other $k-1$ constraints are identical to their classical 
counterparts. They are similar to eq.\ (\ref{Sp_constr})
but the right hand side vanishes.\footnote{These results can
also be understood from symmetry considerations \cite{CSS}.} 
In addition the condition (\ref{Sp_cond}) is modified to
$l+m+\sum_{r=2}^k r\, i_r=2(k-1)-p$, with $p=1,\ldots,k-1$.

For simplicity we focus on $k=2$, or
$Sp(4)$, henceforth. For this case 
one has \cite{CSS}
\bea \label{Sp4_constr}
T_2\, \pf M^0 + {\textstyle\frac12} \pf M^1 &= &2\Lambda^6\ ,\nonumber\\
 \epsilon^{ijkl} M_{ij}^0 M_{kl}^1 &= &0\ ,
\eea
where classically the expression on the left hand side of the first 
equation vanishes. 
Hence, there are  $2\cdot 6 + 1 -2 =11$ physical degrees of freedom
in the effective theory.
Exactly as in the previous $SU(2)$ example
these states can be massive, and 
the superpotential has the generic  form
\bea\label{spfour}
W \hspace{-1em} &= &m_{\hat I} (M,T)\, \hat Q^{\hat I} \hat Q^{\hat I}
                     + Y_{IJK}(M,T)\, Q^I Q^J Q^K  \nonumber\\
&&+ \lambda\ (T_2\, \pf M^0 +{\textstyle\frac12} \pf M^1 -2\Lambda^6)
+\mu\, {\textstyle\frac14}\epsilon^{ijkl} M_{ij}^0 M_{kl}^1\\
&&+\sum_{i<j}\left( m^0_{ij}(M^0_{ij})^2+m^1_{ij}(M^1_{ij})^2\right)
+m T_2^2 + \ldots\ ,  \nonumber
\eea
where $\lambda, \mu$ are the Lagrange multipliers incorporating
the constraints.

Varying with respect to $M^0_{ij},\ M^1_{ij},\ T_2,\ \lambda$ and $\mu$
we obtain 15 equations of motion:
\bea \label{eq_M0}
\half\lambda\ T_2\ \epsilon^{ijkl}M^0_{kl}+\half\mu\ \epsilon^{ijkl}M^1_{kl}
+2m^0_{ij}M^0_{ij} &= &0\quad\forall\ i,j\ ,\\ \label{eq_M1}
\quarter\lambda\ \epsilon^{ijkl}M^1_{kl}+\half\mu\ \epsilon^{ijkl}M^0_{kl}
+2m^1_{ij}M^1_{ij} &= &0\quad\forall\ i,j\ ,\\ \label{eq_T2}
{\textstyle\frac18}\lambda\ \epsilon^{ijkl}M^0_{ij}M^0_{kl}+2mT_2 &= &0\ ,\\
\label{eq_lambda}
T_2\ \epsilon^{ijkl}M^0_{ij}M^0_{kl}+\half\epsilon^{ijkl}M^1_{ij}M^1_{kl}
&= &16\ \Lambda^6\ ,\\ \label{eq_mu}
\epsilon^{ijkl}M^0_{ij}M^1_{kl} &= &0\ .
\eea

As before, we only consider 
solutions for which $\vev W=0$ holds.
With this additional constraint, 
eqs.~(\ref{eq_M0})--(\ref{eq_mu}) imply  $\lambda=\mu=0$.
This can be seen by 
solving  eq.\ (\ref{eq_M0}) for $\mu\ \epsilon^{ijkl}M^1_{kl}$
respectively  eq.\ (\ref{eq_M1}) for $\lambda\ \epsilon^{ijkl}M^1_{kl}$
and inserting the result into (\ref{eq_mu}) multiplied by $\mu$
respectively (\ref{eq_lambda}) multiplied by $\lambda$.
This yields
\bea
-\lambda\ T_2\ \epsilon^{ijkl}M^0_{ij}M^0_{kl}-8\sum_{i<j}m^0_{ij}
(M^0_{ij})^2 &= &0\ ,\\
\lambda\ T_2\ \epsilon^{ijkl}M^0_{ij}M^0_{kl}-\mu\ \epsilon^{ijkl}
M^1_{ij}M^0_{kl}-8\sum_{i<j}m^1_{ij}(M^1_{ij})^2 &= &16\lambda\Lambda^6\ ,
\eea
which together with (\ref{eq_mu}) leads to
\be\label{zwischen}
\sum_{i<j}\left( m^0_{ij}(M^0_{ij})^2+m^1_{ij}(M^1_{ij})^2\right)=
-2\lambda\Lambda^6\ .
\ee
Inserting eqs.~(\ref{zwischen}), (\ref{eq_lambda}),
and (\ref{eq_mu}) into (\ref{spfour})
results in 
\be\label{eq_W}
\vev W=mT_2^2-2\lambda\Lambda^6=0\ .
\ee
All solutions of 
eqs.~(\ref{eq_M0})--(\ref{eq_mu}), (\ref{eq_W}) have $\lambda=0$. 
In addition, for $\lambda=0$ eqs.~(\ref{eq_M0}) and (\ref{eq_M1}) 
imply the following six equations:
\be\label{mucons}
M^0_{kl} (\mu^2-4 m^1_{ij}m^0_{kl}) =0\ \quad
\forall\ i,j,k,l\ {\rm with}\ 
\epsilon^{ijkl}=1\ .
\ee
For generic masses\footnote{This excludes 
special relations among non-vanishing masses.}
 at most one of the
$M^0_{ij}$ can be different from zero
(otherwise $\mu$ would be overdetermined).
The case of one non-vanishing $M^0_{ij}$
is not allowed as a consequence of 
eqs.~(\ref{eq_M0}) and (\ref{eq_mu}), while
the case of all $M^0_{ij}$ vanishing
is not allowed as a consequence of 
eqs.~(\ref{eq_M0}) and (\ref{eq_lambda}).
Thus eq.~(\ref{mucons}) has no solution for 
generic non-vanishing masses.
This implies that at least two of the mass terms
have to vanish which in turn necessarily sets 
$\mu=0$. ´
 
$\lambda=\mu=0$
 considerably simplifies the 
equations of motion. From eqs.\ 
(\ref{eq_M0}), (\ref{eq_M1}), (\ref{eq_T2}) 
we learn that each field  with non-vanishing 
mass term is set to zero. The fields 
with no mass term are only 
constrained by 
eqs.\ (\ref{eq_lambda}), (\ref{eq_mu}). 
Therefore the dimension of
the moduli space is given in general
by the number of 
vanishing  mass terms  minus two.
(However, it is possible that the second constraint
%the two constraints are linearly dependent or one of them 
is trivially satisfied. In this case the dimension of
the moduli space is given by the number of 
vanishing  mass terms  minus one.)

The following examples show the different singularity structures
of the moduli space appearing for different combinations of mass terms.
First, consider the case $m^0_{12}=m^0_{34}=m=0$ which results in a 
two-dimensional moduli space. 
Eqs.~(\ref{eq_M0})--(\ref{eq_mu}) (together
with $\lambda=\mu=0$) lead to 
\bea
M^0_{13}=M^0_{14}=M^0_{23}=M^0_{24} &= &0\ =\ M^1_{ij}\quad 
\forall i,j\ ,\nonumber\\
T_2\, M^0_{12}\, M^0_{34} &= &2\Lambda^6\ .
\eea
A Yukawa coupling of the form
\be
Y_{IJK} \sim T_2+ M^0_{12}+ M^0_{34} + \ldots
\ee
now produces a singularity 
\be
Y_{IJK}(M) \sim 
\frac{2\Lambda^6}{M^0_{12}M^0_{34}} + \ldots 
\ee
as $M^0_{12}\to0$ or $M^0_{34}\to0$.
This is an example for a Yukawa coupling
which depends on two intersecting singular lines.

As a second example, consider
the case  $m^1_{12}=m^1_{34}=0$ which corresponds
to a one-dimensional moduli space.
Eqs.~(\ref{eq_M0})--(\ref{eq_mu}) (together
with $\lambda=\mu=0$) imply
\bea
T_2=M^1_{13}=M^1_{14}=M^1_{23}=M^1_{24} &= &0\ =\ M^0_{ij}\quad 
\forall i,j\ ,\nonumber\\
M^1_{12}\, M^1_{34} &= &4\Lambda^6\ .
\eea
These constraints lead to the same singularity 
structure (cf.\  eqs.\ 
(\ref{sol_su2}), (\ref{singularity})) as 
in the $SU(2)$ case discussed in section~2.

A different situation occurs
for $m^0_{12}=m^0_{34}=m^0_{13}=m^1_{24}=m^1_{34}=m=0$,
since eq.~(\ref{eq_mu}) is not automatically satisfied. 
The moduli space has dimension four and is constrained by
\bea\label{sol2}
M^{0,1}_{14}=M^{0,1}_{23}=M^0_{24}
=M^1_{12}=M^1_{13} &= &0\ ,\nonumber\\
T_2\, M^0_{12}\, M^0_{34} &= &2\Lambda^6\ ,\\
M^0_{12}\, M^1_{34}-M^0_{13}\, M^1_{24} &= &0\ .\nonumber
\eea
If we solve the last two equations for $T_2$ and 
$M^0_{12}$, say, we find for a Yukawa coupling
of the form
\be
Y(M) \sim
T_2+M^0_{12}+M^0_{34}+M^0_{13}+M^1_{24}+M^1_{34}+\ldots
\ee
the singularities
\be
Y(M) \sim {2M^1_{34}\Lambda^6\over M^0_{34}M^0_{13}M^1_{24}}
+{M^0_{13} M^1_{24}\over M^1_{34}}+\ldots\ ,
\ee
as $M^0_{34},\ M^0_{13},\ M^1_{24}$, or $M^1_{34}\to0$. 
There are four singular components but only three of them appear 
in the same term of the Yukawa couplings. In general, we will
call $l$ singular components {\em intersecting} if they give
rise to a singularity of degree $l$ in the Yukawa couplings
when going to zero simultaneously. One finds that at most
three intersecting singular components can be generated in 
the $Sp(4)$ model.

Before we turn to the discussion of the general
situation in $Sp(2k)$, let us note that 
the $SU(2)$ gauge theory discussed in section~2
arises on the Higgs branch of the 
$Sp(4)$ gauge theory considered above. 
An expectation value of the antisymmetric tensor 
$A^{\alpha\beta}$ breaks $Sp(4)\to SU(2)\times SU(2)$,
and one recovers two copies of the $SU(2)$ model of the previous
section. This follows directly from ref.~\cite{CSS}, where it 
was shown that the confining superpotential of the $Sp(2k)$ 
theory breaks into a sum of $SU(2)$-type superpotential terms 
when a VEV is given to the antisymmetric tensor.
The D-flatness condition implies (up to gauge rotations)
\be
\vev{A^{\alpha\beta}}\ =\ \left(
        \begin{array}{cccc}
           0 &v &0 &0\\
           -v &0 &0 &0\\
           0 &0 &0 &-v\\
           0 &0 &v &0
        \end{array}\right) \ ,
\ee
where $\half v^2=T_2\equiv{\textstyle\frac18}\vev{A_{\alpha}^{\ \beta}}
\vev{A_{\beta}^{\ \alpha}}$.
Inserted into eqs.~(\ref{MTdef}), one has
\bea
M^0_{ij}=q_i^\alpha\epsilon_{\alpha\beta}q_j^\beta
        +q_i^{\alpha+2}\epsilon_{\alpha\beta}q_j^{\beta+2}
   &\equiv &\hat M^{(1)}_{ij}+\hat M^{(2)}_{ij}\ ,\\
M^1_{ij}=-vq_i^\alpha\epsilon_{\alpha\beta}q_j^\beta 
        +vq_i^{\alpha+2}\epsilon_{\alpha\beta}q_j^{\beta+2}
   &\equiv &-v\hat M^{(1)}_{ij}+v\hat M^{(2)}_{ij}\ .
\eea
The two constraints (\ref{eq_lambda}) and 
(\ref{eq_mu}) written in terms of  
$\hat M^{(1)},\hat M^{(2)}$
read
\bea
v^2\pf \hat M^{(1)} + v^2\pf \hat M^{(2)} &= &2\Lambda^6_{Sp(4)}\ ,\\
-v\pf \hat M^{(1)} + v\pf \hat M^{(2)} &= &0\ .
\eea
Using the scale matching condition \cite{CSS}
$\Lambda^6_{Sp(4)}=v^2\Lambda^4_{SU(2)}$,
we see that we have two copies of the $SU(2)$ moduli 
space with the
correct constraints (\ref{su2_constr})
\be
\pf \hat M^{(1)}=\Lambda^4_{SU(2)}\ ,\quad 
\pf \hat M^{(2)}=\Lambda^4_{SU(2)}\ .
\ee

So far we focused on a non-perturbative gauge group
$\GNP=Sp(4)$ and saw that up to four singular components 
in the Yukawa couplings can be generated. 
More singular components arise in an $Sp(2k)$ gauge 
theory, once we have $k>2$. The main point of the mechanism
that generates singularities in the Yukawa couplings is the
observation that a quantum constraint for a product of
$l+1$ different moduli fields gives rise to $l$ singular
components when inserted in generic Yukawa couplings.
We first restrict ourselves to the case where all singular
components are intersecting in the sense defined above
and where their number is equal to the dimension of the moduli
space. As a consequence, only one of the $k$ constraints on
the quantum moduli space is relevant, namely the one of the
form (\ref{Sp_constr}). The $k-1$ homogeneous constraints,
for which the right hand side vanishes, must be trivially
satisfied; else they would either involve non-singular
moduli fields or produce additional singular components
which do not intersect the others.
 
We are interested in the smallest \nonpert $Sp(2k)$ gauge group
that is able to generate $l$ intersecting singular components
in the Yukawa couplings.
Because of (\ref{Sp_constr}) and (\ref{Sp_cond}) we need to look
at a term of the form
\be
T_2\, T_3 \ldots T_l\, \pf M^0
\ee
to find the minimal $k$ required for a quantum constraint for $l+1$
different moduli fields. The condition (\ref{Sp_cond}) now reads
$l+m+\sum ri_r=\sum_{r=2}^lr=\half l(l+1)-1\ =\ 2(k-1)$, i.e., 
the considered term is allowed if $4k=l(l+1)+2$. As $k$ is integer
there is not a solution for each $l$, but whenever
\be \label{cond_l}
4k\ge l(l+1)+2\ ,
\ee
a term with at least $l+1$ different factors in the constraint
(\ref{Sp_constr}) is possible. Inserted into generic Yukawa couplings
this yields a singularity of degree $l$.
Unfortunately, the coefficients of the sum (\ref{Sp_constr})
are not known for general $k$; there are terms which are allowed by the 
symmetries but are not generated by the gauge dynamics because their 
coefficients vanish. Therefore the condition (\ref{cond_l}) is only
necessary but not sufficient for the appearance of the terms considered
above. (Up to $k=4$ the coefficients are known \cite{CSS}, and the
condition (\ref{cond_l}) is also sufficient.)

If the dimension of the moduli space is greater than the number of
intersecting singular components, then the condition (\ref{cond_l})
to get $l$ intersecting singular components is no longer valid. 
This is because the $k-1$ homogeneous constraints now play a role
in that they can lead to more intersecting singular components,
as we have seen in our last example of the $Sp(4)$ model. In this
example we found $l=3$ intersecting singular components. 
The solutions of the constraint equations for $Sp(2k)$ get more
and more complex with increasing $k$. The case $k=3$ is still
tractable, and it turns out that one can have $l=5$. 

%%%%%%%%%%%%%%%%%%%%%%%%%%%%%%%%%%%%%%%%%%
\section{Heterotic vacua with non-perturbative
gauge group $U(2k)$}
So far we discussed heterotic vacua in $d=4$ which
are related to heterotic $SO(32)$ vacua in $d=6$ 
with small instantons sitting on smooth points
of the $K3$.
The situation is more involved when an $SO(32)$
instanton shrinks on a singularity of the
underlying CFT or K3 manifold.
(For simplicity we only discuss the case of
an $A_1$ singularity.)
In this case one has to distinguish between instantons
`with vector structure' and  
`without vector structure' \cite{bele}.\footnote{
This terminology refers to properties of the $SO(32)$
connection at infinity. See refs.~\cite{bele,aspinwall}
for details.}  
For instantons without vector structure
the instanton moduli space is conjectured to be
isomorphic to the Higgs branch of a
$U(2k)$ gauge theory with 16 hypermultiplets
in the fundamental ($\fun$) $2k$-dimensional representation and
two antisymmetric tensors ($\as$) in the
$k(2k-1)$-dimensional representation \cite{bele,DM}.
Furthermore, the $U(1)$ factor is spontaneously
broken by the Green-Schwarz mechanism and the the low energy
effective theory only has $SU(2k)$ massless gauge boson \cite{bele}. 

Small instantons with vector structure on an $A_1$ singularity show a
more complicated behaviour \cite{aspinwall,intriligator,AM,DM,BI}.
The case where less than four instantons
coalesce on the singularity  is 
not fully understood yet,
while the moduli space of four (and more)
instantons on the singularity
has a Higgs branch
and a Coulomb branch. On the Coulomb branch the dimension of the 
moduli space has been reduced by 29, but an additional tensor 
multiplet is present. 

Here we restrict ourselves to $k$
instantons without vector structure shrinking
on an $A_1$ singularity of the $K3$ fibre.
Again the computation of the spectrum in $d=4$ 
can be performed following the methods 
of ref.~\cite{KSS}. 
One finds two flavours of fundamentals, $q_i, \bar q_i, i=1,2$,  
as well as one flavour
of antisymmetric tensors $A, \bar A$.\footnote{By 
giving an appropriate vacuum expectation value to 
one of the two antisymmetric tensors of $SU(2k)$ one 
arrives at the $Sp(2k)$ gauge theory with the precise
spectrum discussed in section~3.}
As before, this is an asymptotically free gauge 
theory with 
$c_{SU(2k)}=2k-4\cdot\frac12-2(k-1)=0$ 
for any $k$.
Consequently, this theory confines 
below  $\Lambda_{SU(2k)}$, but no
non-perturbative superpotential is generated
by the strong coupling dynamics.

For simplicity, we first concentrate on $\GNP=SU(4)$ where 
$A\cong\bar A$ holds. 
Therefore the two antisymmetric 
tensors $A_r,\ r=1,2,$ transform as a doublet under an 
additional $SU(2)$ flavour symmetry. The low energy degrees 
of freedom are found to be \cite{confine}
\bea
M^0_{ij} &:= &q_i\bar q_j\ =\ q_i^\alpha\bar q_{j\alpha}\ ,
\quad i,j=1,2\ ,\nonumber\\
M^1_{ij} &:= &q_iA^2\bar q_j\ =\ \quarter\epsilon_{\alpha\beta\gamma\delta}
\epsilon^{rs}q_i^\alpha A_r^{\beta\gamma}A_s^{\delta\lambda}\bar 
q_{j\lambda}\ ,\nonumber\\
H_r &:= &q_iA_rq_j\ =\ \quarter\epsilon_{\alpha\beta\gamma\delta}\epsilon^{ij}
q_i^\alpha A_r^{\beta\gamma} q_j^\delta\ ,\quad r=1,2\ , \\
\bar H_r &:= &\bar q_i A_r \bar q_j\ =\ \half\epsilon^{ij}\bar q_{i\alpha} 
A_r^{\alpha\beta}\bar q_{j\beta}\ ,\nonumber\\
T_{rs} &:= &A_rA_s\ =\ {\textstyle\frac18}\epsilon_{\alpha\beta\gamma\delta}
A_r^{\alpha\beta}A_s^{\gamma\delta}\ .\nonumber
\eea
These singlet fields satisfy the additional constraints
\bea \label{su4_constr}
\det T\det M^0 - \epsilon^{rs}\epsilon^{tu}T_{rt}H_s\bar H_u - \det M^1
&= &\Lambda^8, \nonumber\\
\epsilon^{ij}\epsilon^{kl}M^0_{ik}M^1_{jl}+\epsilon^{rs}H_r\bar H_s 
&= &0\ .
\eea
Implementing these constraints in the superpotential via
Lagrange multipliers $\lambda$ and $\mu$, we have
\bea
W &= &m_{\hat I} (M,H,T)\, \hat Q^{\hat I} \hat Q^{\hat I}
      + Y_{IJK}(M,H,T)\, Q^I Q^J Q^K  \nonumber\\
&&+ \lambda\ (\det T\det M^0 - TH\bar H - \det M^1 -\Lambda^8)
+\mu\ (M^0 M^1+H\bar H)\\
&&+\sum_{i,j}\left( m^{(0)}_{ij}(M^0_{ij})^2+m^{(1)}_{ij}(M^1_{ij})^2\right)
+\sum_{r\le s}m^{(2)}_{rs} (T_{rs})^2 
+\sum_r \left( m^{(3)}_r (H_r)^2+m^{(4)}_r (\bar H_r)^2 \right)\nonumber\ \\
&&+ \ldots\ .  \nonumber
\eea
Using the equations of motion, it is easy to show that, in direct analogy
to the $Sp(4)$ case,
\be
\vev W=\sum_{r\le s}m^{(2)}_{rs} (T_{rs})^2-\lambda\Lambda^8\ .
\ee
As for $Sp(4)$, the superpotential vanishes 
at the minimum for $\lambda=0$.
However, in this case we were not able to show
that this condition is also sufficient.
Nevertheless we continue to consider only 
solutions with $\lambda=0$.
As before, $\lambda=0$ implies $\mu=0$ for 
generic mass terms, and the equations of motion 
can be easily solved. All fields 
with non-vanishing mass terms are set to zero,
while the remaining degrees of freedom must 
satisfy the constraints (\ref{su4_constr}).

The structure of the singularities again depends
on the mass terms  for the confined degrees of freedom.
For example, mass terms for the fields $M^1_{ij}, H_r, 
\bar H_r, M^0_{12}, M^0_{21}$ and $T_{12}\ (i,j,r=1,2)$ 
result in a three-dimensional moduli space described by
\bea
M^1_{ij}=H_r= \bar H_r =M^0_{12} = M^0_{21} =T_{12} &= &0\ ,\nonumber\\
T_{11}T_{22}M^0_{11}M^0_{22} &= &\Lambda^8\ .
\eea
Inserting this into generic Yukawa couplings results in a
singular locus with three intersecting components:
\be
Y(M)\sim {\Lambda^8\over T_{22}M^0_{11}M^0_{22}}+\ldots\ .
\ee

A more complicated structure of the moduli space is obtained
if only $M^1_{12}$, $M^1_{21}$, $M^1_{22}$, $M^0_{12}$, $M^0_{21}$, 
$H_1$, $\bar H_2$ and $T_{12}$ obtain mass terms. Solving the
equations of motion, one finds a five-dimensional moduli space:
\bea
M^1_{12}=M^1_{21}=M^1_{22}=M^0_{12}=M^0_{21}=H_1=\bar H_2=T_{12} &= &0\ ,
\nonumber\\
T_{11}T_{22}M^0_{11}M^0_{22} &= &\Lambda^8\ ,\\
M^0_{22}M^1_{11}-H_2\bar H_1 &= &0\nonumber
\eea
If we eliminate $T_{11}$ and $M^0_{22}$ by these two constraints,
a generic Yukawa coupling will have singularities of the form
\be
Y(M)\sim {M^1_{11}\Lambda^8\over T_{22}M^0_{11}H_2\bar H_1}
        +{H_2\bar H_1\over M^1_{11}}+\ldots\ .
\ee
There are five singular components four of which are intersecting
in that they appear in the same coupling term.

In the general case of $\GNP=SU(2k),\ k>2,$ the situation is more complex.
The antisymmetric tensor $A^{\alpha\beta}$ has now to be distinguished
from its conjugate $\bar A_{\alpha\beta}$. Therefore the confined 
spectrum differs from the $SU(4)$ case. The low energy degrees of
freedom are\footnote{Note that the `mesons' $M,P,\bar P$ are
identical to the meson singlets of the dual magnetic theory found
in ref.\ \cite{ILS}.} \cite{confine}
\bea
M^l_{ij} &:= &q_i(\bar AA)^l\bar q_j\ =\ 
q_i^\alpha \bar A_{\alpha\beta_1}A^{\beta_1\beta_2}\cdots
A^{\beta_{2l-1}\gamma}\bar q_{j\gamma}\ ,
\quad i,j=1,2,\ l=0,\ldots,k-1\ ,\nonumber\\
P^m &:= &q(\bar AA)^m\bar Aq\ =\ 
\epsilon^{ij}q_i^\alpha \bar A_{\alpha\beta_1}A^{\beta_1\beta_2}\cdots
\bar A_{\beta_{2m}\gamma}q_j^\gamma\ ,
\quad m=0,\ldots,k-2\ ,\nonumber\\
\bar P^m &:= &\bar qA(\bar AA)^m\bar q\ =\ 
\epsilon^{ij}\bar q_{i\alpha} A^{\alpha\beta_1}\bar A_{\beta_1\beta_2}\cdots
A^{\beta_{2m}\gamma}q_{j\gamma}\ ,\nonumber\\
B_0 &:= &A^k\ =\ \epsilon_{\alpha_1\cdots\alpha_{2k}}A^{\alpha_1\alpha_2}
\cdots A^{\alpha_{2k-1}\alpha_{2k}}\ ,\\
\bar B_0 &:= &\bar A^k\ =\ \epsilon^{\alpha_1\cdots\alpha_{2k}}
\bar A_{\alpha_1\alpha_2} \cdots \bar A_{\alpha_{2k-1}\alpha_{2k}}\ ,
\nonumber\\
B_2 &:= &A^{k-1}qq\ =\ \epsilon_{\alpha_1\cdots\alpha_{2k}}A^{\alpha_1\alpha_2}
\cdots A^{\alpha_{2k-3}\alpha_{2k-2}}
q^{\alpha_{2k-1}}q^{\alpha_{2k}}\ ,\nonumber\\
\bar B_2 &:= &\bar A^{k-1}\bar q\bar q\ =\ \epsilon^{\alpha_1\cdots\alpha_{2k}}
\bar A_{\alpha_1\alpha_2} \cdots \bar A_{\alpha_{2k-3}\alpha_{2k-2}}
\bar q_{\alpha_{2k-1}}\bar q_{\alpha_{2k}}\ ,\nonumber\\
T_n &:= &(\bar AA)^n\ =\ \bar A_{\alpha\beta_1}A^{\beta_1\beta_2}\cdots
A^{\beta_{2n-1}\alpha}\ ,\quad n=1,\ldots,k-1\ .\nonumber
\eea

These fields are not all independent but obey constraint equations which, in
general, are quite involved but can be derived from the confining
superpotential of the model with three quark flavours ($N_f=3$) \cite{confine}.
By the gauge and global symmetries this superpotential is forced to be
of the form
\bea
W_{N_f=3} &= &{1\over\Lambda^{4k-1}}\ \Bigg(
        \sum_{\{i_r\},j,m,n,\atop l_1,l_2,l_3}\!\!\!\!\!c_{ijlmn}
        \prod_{r=1}^{k-1} (T_r)^{i_r} (B_0\bar B_0)^j
        (M^{l_1}M^{l_2}M^{l_3}+\alpha_{mn} M^mP^n\bar P^n)
        \nonumber\\
  &&\hspace{2cm}+\beta M^0B_2\bar B_2\Bigg)\ ,
\eea
where flavour indices have been suppressed, and the sum goes over all 
integers 
\bea
&0\le l_1,l_2,l_3,m\le k-1\ ,\quad 0\le n\le k-2\ ,\quad
   i_r\ge 0\ ,\quad j=0,1\ ,\nonumber\\
&\displaystyle{\rm with}\quad
 \left.\begin{array}{c} l_1+l_2+l_3\\ m+2n+1\end{array}
 \right\}+kj+\sum_{r=1}^{k-1}ri_r=2(k-1)\ .
\eea
Again, the coefficients can be calculated by requiring that the equations
of motion reproduce the classical constraints.
The quantum constraints on the $N_f=2$ moduli space are obtained by giving 
a large mass to one quark flavour and integrating out the massive modes.
Only one of these constraints differs from the corresponding classical 
constraint. It has the form
\be \label{SU_constr}
\sum_{\{i_r\},j,\atop l,m,n} \tilde c_{ijlmn}
          \prod_{r=1}^{k-1} (T_r)^{i_r} (B_0\bar B_0)^j
          (M^{l}M^{m}+\tilde\alpha_n P^{n}\bar P^{n})
           \ +\ \tilde\beta B_2\bar B_2=\Lambda^{4k}\ .
\ee
where it is summed over all $i_r,j,l,m,n$ that satisfy
\be \label{SU_cond}
\left.\begin{array}{c} l+m\\ 2n+1\end{array}
\right\}+kj+\sum_{r=1}^{k-1}ri_r=2(k-1)\ .
\ee

To see how many intersecting singular components are possible for general 
$k$, we look again at products of $l+1$ moduli fields in eq.\ 
(\ref{SU_constr}).\footnote{As in the $Sp(2k)$ case, the following
argument is restricted to the case where the dimension of the moduli
space equals the number of intersecting singular components.}
Relevant terms with $l+1$ factors which are allowed for minimal $k$ 
are of the form
\bea
T_1\cdots T_{l-1}M^0M^0\nonumber\quad {\rm or}\\
T_1\cdots T_{l-3}B_0\bar B_0M^0M^0\nonumber\ .
\eea
By eq.\ (\ref{SU_cond}), the term appearing in the first line is 
allowed if $4k=l(l-1)+4$, while the second one requires $2k=(l-2)(l-3)+4$. 
One finds that for $l$ intersecting singular components to be possible,
necessarily
\bea
2k \ge (l-2)(l-3)+4 &{\rm \ \ if\ \ } &l\le 6\ ,\\
4k \ge l(l-1)+4 &{\rm \ \ if\ \ } &l\ge 7\ .
\eea

Let us summarize. In this paper we expanded
on a mechanism suggested in ref.~\cite{KSS}
which gives rise to singular couplings
in the low energy effective theory of 
$N=1$ heterotic string vacua.
In addition to the perturbative gauge group of
a string vacuum, also a confined
non-perturbative gauge group can be present
whose strong coupling dynamics renders the
Yukawa and gauge couplings singular.
This phenomenon occurs for asymptotically free gauge theories
with the additional property that $c$, defined
in eq.~(\ref{cdef}), vanishes. 
Such non-perturbative
gauge theories do appear
in $K3$ fibred Calabi--Yau
compactifications and are directly related
to non-perturbative gauge theories
in $d=6$.
The structure of the singularities are qualitatively
similar to the singularities in Calabi--Yau
compactifications, but a more detailed quantitative
comparison is still lacking. 
It would also be interesting
to extend our work for other 
classes of heterotic vacua (see for example ref.~\cite{CS})
as well as
to study the possibility
of extremal transitions between $N=1$ heterotic
string vacua along the lines of refs.~\cite{GMS,K-S}.

\vspace{1cm}

\noindent {\bf Acknowledgement}

We have greatly benefited from discussions
and email correspondence with  P.\ Aspinwall,
T.~Banks, S.~F\"orste, K.~Intriligator, 
S.~Kachru, V.~Kaplunovsky,
T.~Mohaupt, G.~Moore, R.~Schimmrigk, 
M.~Schmaltz, E.~Silverstein, 
J.~Sonnenschein, and S.~Theisen.

The work of M.K. is supported by the DFG. 
The work of J.L. is supported in part by GIF -- the German--Israeli
Foundation for Scientific Research
and NATO under grant CRG~931380.


\begin{thebibliography}{150}
%
\bibitem{schmal}E.\ Witten, 
{\it ``Small instantons in string theory''},
Nucl.\ Phys.\ {\bf B460} (1996) 541, hep-th/9511030.
\label{schmal}
%
\bibitem{MV} D.R.\ Morrison and C.\ Vafa, 
{\it ``Compactification of F-Theory on Calabi--Yau Threefolds''}, 
Nucl.\ Phys.\ {\bf B473} (1996) 122,
hep-th/9602114; Nucl.\ Phys.\ {\bf B476} (1996) 437,
hep-th/9603161.
%
\bibitem{hanany}O.J.\ Ganor and A.\ Hanany, 
{\it ``Small $E_8$ instantons
and tensionless non critical strings''}, 
Nucl.\ Phys.\ {\bf B474} (1996) 122,
hep-th/9602120.
%
\bibitem{SW}N.\ Seiberg and E.\ Witten, 
{\it ``Comments on string dynamics in six dimensions''}, 
Nucl.\ Phys.\ {\bf B471} (1996) 121, hep-th/9603003.
%
\bibitem{bele}M.\ Berkooz, R.\ Leigh, J.\ Polchinski,
J.\ Schwarz, N.\ Seiberg, and E.\ Witten, 
{\it ``Anomalies, Dualities, and Topology
of $D=6 \, N=1$ superstring vacua''}, 
Nucl.\ Phys.\ {\bf B475} (1996) 115, 
hep-th/9605184.
\label{bele}
%
\bibitem{aspinwall} P.S.\ Aspinwall,
{\it ``Point-like Instantons and the $Spin(32)/Z_2$
 Heterotic String''},
hep-th/9612108.
%
\bibitem{intriligator} K.\ Intriligator,
{\it ``RG Fixed Points in Six Dimensions 
via Branes at Orbifold Singularities''}, 
hep-th/9702038.
%
\bibitem{AM} P.S.\ Aspinwall and D.\ Morrison,
{\it ``Point-like Instantons on $K3$ orbifolds''},
hep-th/9705104.\label{AM}
%
\bibitem{strominger} A.\ Strominger,
{\it ``Massless black holes and conifolds in string theory''},
Nucl.\ Phys.\ {\bf B451} (1995),
hep-th/9504090.
%
\bibitem{various} E.~Witten, 
{\it ``String theory dynamics in various dimensions''}, 
Nucl.\ Phys.\ {\bf B443} (1995) 85, hep-th/9503212.
%
\bibitem{KSS} S.\ Kachru, N.\ Seiberg, and E.\ Silverstein,
{\it ``SUSY Gauge Dynamics and Singularities of 
4d N=1 String Vacua''},
Nucl.\ Phys.\ {\bf B480} (1996) 170,
hep-th/9605036.
%
\bibitem{BS} M.\ Bianchi and A.\ Sagnotti,
{\it ``On The Systematics of Open String Theories''},
Phys.\ Lett.\ {\bf 247B} (1990) 517;
{\it ``Twist Symmetry and Open String Wilson Lines''},
Nucl.\ Phys.\ {\bf B361} (1991) 827.
\label{BS}
%
\bibitem{GP} E.\ Gimon and  J.\ Polchinski,
{\it ``Consistency Conditions for Orientifolds and D-Manifolds''}, 
Phys.\ Rev.\ {\bf D54} (1996) 1667, 
hep-th/9601038. 
%
\bibitem{DM} M.\ Douglas and G.\ Moore,
{\it ``D-branes, Quivers, and ALE Instantons''},
hep-th/9603167.\label{DM}
%
\bibitem{klein} M.\ Klein and J.\ Louis,
{\it ``Singularities in $d=4$,  $N=1$ Heterotic String Vacua''},
hep-th/9707047.
%
\bibitem{LF} For a recent review see for example,
K.\ F\"orger and J.\ Louis, 
{\it ``Holomorphic Couplings in String Theory''},
Nucl.\ Phys.\ B (Proc.\ Suppl.)
{\bf 55B} (1997) 33,  hep-th/9611184.
%
\bibitem{KS} S.\ Kachru and E.\ Silverstein,
{\it ``Singularities, Gauge Dynamics, 
and Nonperturbative Superpotentials in String Theory''},
Nucl.\ Phys.\ {\bf B482} (1996) 92, 
hep-th/9608194;\\
S.\ Kachru, {\it ``Aspects of N=1 String Dynamics''},
hep-th/9705173.
%
\bibitem{CDGP}
P.\ Candelas, X.\ De la Ossa, P.S.\ Green, and
L.\  Parkes, 
{\it ``An Exactly Soluble Superconformal Field theory
from a Mirror Pair of Calabi--Yau Manifolds''},
Nucl.\ Phys.\ {\bf B258} (1991) 118.
%
\bibitem{BCOV}
M.~Bershadsky, S.~Cecotti, H.~Ooguri, and C.~Vafa,
 {\it ``Holomorphic Anomalies in Topological Field Theories''},
Nucl.\ Phys.\ {\bf B 405} (1993) 279, hep-th/9302103;\\
 {\it ``Kodaira-Spencer Theory of Gravity 
and Exact Results for Quantum String Amplitudes''},
Comm.\ Math.\ Phys.\ {\bf 165} (1994) 311, 
hep-th/9309140.
%
\bibitem{K-S} S.\ Kachru and E.\ Silverstein,
{\it ``Chirality Changing Phase Transitions in 4d String Vacua''},
hep-th/9704185.
%
\bibitem{ibanez} G.\ Aldazabal,  A.\ Font,  
L.E.\ Ibanez,  A.M.\ Uranga, and  G.\ Violero,
{\it ``Non-Perturbative Heterotic $D=6,4$
 Orbifold Vacua''},
hep-th/9706158.
%
\bibitem{VW} C.\ Vafa and E.\ Witten,
{\it ``Dual String Pairs With $N=1$ And $N=2$
 Supersymmetry In Four Dimensions''}, 
hep-th/9507050.
%
\bibitem{CDFKM}
P.\ Candelas, X.\ De la Ossa, A.\ Font,
S.\ Katz, and D.\  Morrison, 
{\it ``Mirror Symmetry for Two Parameter Models -- I''},
Nucl.\ Phys.\ {\bf B416} (1994) 481, hep-th/9308083.
%
\bibitem{KLM} A.\ Klemm, W.\ Lerche, and P.\ Mayr,
{\it ``$K3$ Fibrations and Heterotic-Type II 
String Duality''},
Phys.\ Lett.\ {\bf B357} (1995) 313, 
hep-th/9506112.
%
\bibitem{ADS} I.\ Affleck, M.\ Dine, and N.\ Seiberg,
{\it ``Dynamical Supersymmetry Breaking in Four Dimensions
And Its Phenomenological Implications''}, 
Nucl.\ Phys.\ {\bf B256} (1985) 557.
%
\bibitem{confine} C.\ Cs\'aki, M.\ Schmaltz, and W.\ Skiba,
{\it ``A Systematic Approach to Confinement in N=1 Supersymmetric Gauge
Theories''}, Phys.\ Rev.\ Lett.\
{\bf 78} (1997) 799, hep-th/9610139;\\
{\it ``Confinement in $N=1$ SUSY Gauge Theories and Model 
Building Tools''},
Phys.\ Rev.\ {\bf D55} (1997) 7840,
hep-th/9612207.
%
\bibitem{Seib_D49} N.\ Seiberg, 
{\it ``Exact Results on the space of vacua
of four-dimensional SUSY gauge theories''}, 
Phys.\ Rev.\ {\bf 49} (1994) 6857, hep-th/9402044.
%
\bibitem{sagnotti} A.\ Sagnotti,
{\it ``A Note on the Green--Schwarz Mechanism in Open String Theory''},
Phys.\ Lett.\ {\bf 294B} (1992) 196, 
hep-th/9210127.
%
\bibitem{DMW}M.J.\ Duff, R.\ Minasian, and E.\ Witten,
{\it ``Evidence for heterotic/heterotic duality''},
Nucl.\ Phys.\ {\bf B465} (1996) 413, hep-th/9601036.
%
\bibitem{KL}
V.\ Kaplunovsky and J.\ Louis,
{\it ``On Gauge Couplings in String Theory''},
Nucl.\ Phys.\ {\bf B444} (1995) 191,
hep-th/9502077.
%
\bibitem{CSS} C.\ Cs\'aki, M.\ Schmaltz, and W.\ Skiba,
{\it ``Exact Results and Duality for SP(2N) SUSY 
Gauge Theories with an Antisymmetric Tensor''},
Nucl.\ Phys.\ {\bf B487} (1997) 128, 
hep-th/9607210.
%
\bibitem{CK} P. Cho and P. Kraus, 
{\it ``Symplectic SUSY Gauge Theories with Antisymmetric Matter''},
Phys.\ Rev.\ {\bf D54} (1996) 7640,
hep-th/9607200.
%
\bibitem{IP} K.\ Intriligator and P.\ Pouliot,
{\it ``Exact Superpotentials, Quantum Vacua and Duality 
in Supersymmetric $SP(N_c)$ Gauge Theories''},
Phys.\ Lett.\ {\bf B353} (1995) 471, hep-th/9505006;\\
K.\ Intriligator
{\it ``New RG Fixed Points and Duality in Supersymmetric 
$SP(N_c)$ and $SO(N_c)$ Gauge Theories''},
Nucl.\ Phys.\ {\bf B448} (1995) 187, hep-th/9505051.
%
\bibitem{BI} J.\ Blum and K.\ Intriligator,
{\it ``Consistency Conditions for Branes at Orbifold Singularities''},
hep-th/9705030;
{\it ``New Phases of String Theory and 6d RG Fixed Points via Branes at Orbifold Singularities''},
hep-th/9705044.
%
\bibitem{ILS} K.\ Intriligator, R.G.\ Leigh, and M.J.\ Strassler,
{\it ``New Examples of Duality in Chiral and Non-Chiral
Supersymmetric Gauge Theories''},
Nucl.\ Phys.\ {\bf B456} (1995) 567, 
hep-th/9506148.
%
\bibitem{CS} P.\ Candelas and H.\ Skarke,
{\it ``F-theory, SO(32) and Toric Geometry''},
hep-th/9707049.
%
\bibitem{GMS} B.R.\ Greene, D.R.\ Morrison, and A.\ Strominger,
{\it ``Black Hole Condensation and the Unification of String Vacua ''},
Nucl.\ Phys.\ {\bf B451} (1995) 109, 
hep-th/9504145.
%
\end{thebibliography}
\end{document}